\documentclass[prb,twocolumn,aps,floatfix]{revtex4}

\usepackage{graphicx}
\usepackage{color}
\usepackage{bm}
\usepackage{amssymb}
\usepackage{amsmath}
\usepackage{subfigure}
\usepackage{tikz}
\usetikzlibrary{shapes,arrows}
\usepackage{amsmath}
\usepackage{appendix}
\usepackage{multirow}
\usepackage{booktabs}

\begin{document}

\title{Spin wave theory for interacting hardcore bosons on cubic lattices: a comparative study}
\author{Vipin Kerala Varma and Hartmut Monien}
\affiliation{Bethe Center for Theoretical Physics, Universit\"{a}t Bonn, Germany\\}
\date{\today}
\begin{abstract}
Motivated by excellent comparisons of spin wave theory (SWT) with Quantum Monte Carlo (QMC) calculations of non-interacting hardore bosons (Bernardet et al.) and superlattices (Hen et al.), we develop an SWT for interacting hardcore bosons on a $d$-dimensional cubic lattice at zero temperature and compare thermodynamical properties with results of Stochastic Series Expansions (SSE): there is good comparison between SWT and SSE results for the simple cubic lattice at small repulsive strengths but differences arise between the results of the two approaches for the square lattice at strong repulsions. The boundaries of the three phases - Mott insulating (MI), superfluid (SF) and N\'{e}el solid (NS) - can be readily estimated from calculations of particle density, signalling either a first order (SF $\leftrightarrow$ NS) or second order transition (MI $\leftrightarrow$ SF). Comparison is made with results of the Bethe ansatz in one dimension and QMC in two dimensions.
\end{abstract}

\maketitle

Bosons on a lattice appear as a natural model for several physical systems. Such a model was used to elucidate the $\lambda$-transition in liquid helium II\cite{Matsubara} which, at the time, was still lacking a microscopic basis. Furthermore, the authors in Ref. [1] showed that the partition function of a suitably approximated model - which later came to be variously described as the hardcore boson Hubbard model (HCB) -  is equal to the partition function of a system of vector spins - the Heisenberg model in a longitudinal magnetic field - as defined on an $infinite$ lattice, when a suitable identification of constants and operators between the two models is made. The latter system's partition function, they argued, could be calculated by an approximation scheme like spin-wave analysis. Moreover, an additional reason for preferably treating the corresponding spin model is one of calculational convenience vis-\'{a}-vis keeping track of the unusual commutation relation of hardcore bosons (see next section). \\
Within this approximation, several studies at zero and finite temperatures\cite{Matsuda, Liu, Scalettar} were undertaken. Although Ref. [2] focussed on describing the phase boundaries, many thermodynamical quantities were not explicitly calculated. The authors of Ref. [3] repeated the calculations, on a square lattice mostly near half-filling, in order to resolve some previously conflicting results regarding the spectrum in the earlier predicted phases. The focus of the present work, unlike previous reports, will not be on the existence or otherwise of a supersolid phase but rather on calculating thermodynamic quantities using a collinear spin wave theory (SWT) in the interacting model and comparing the results with other methods.\\
In fact, with the subsequent development of sophisticated Quantum Monte Carlo (QMC) techniques, SW treatment has been seen to provide a surprisingly adequate description for many physical observables in the HCB model like particle density, ground state energy and suchlike for (a) the non-interacting case\cite{Bernardet} and (b)  superlattice systems\cite{Hen}. Indeed, as Bernardet et al. and Hen et  al. showed, the discrepancies, if any, between the results of the two approaches - QMC and SWT - are extremely small for many thermodynamical properties. This provides further incentive to revisit the corresponding nearest neighbour (NN) interacting problem for a general cubic lattice using SWT. \\
The goal of the paper is to provide an SW treatment for the interacting HCB model on a $d$-dimensional cubic lattice, calculate certain thermodynamic quantities of interest and compare our results with some known solutions, where available, in any particular dimension. In fact, it was suggested\cite{Bernardet} that SWT should be sufficiently good even in the interacting case; we find that, starting from a collinear phase, the comparison between SWT and results of Stochastic Series Expansion (SSE) is reasonable but not as good as for the non-interacting case for the square lattice with strong repulsions $V$. For the simple cubic lattice, there is good agreement between the two approaches at a lower $V$ value.

\section{Model}
The hardcore boson model with nearest neighbour interaction on a $d$-dimensional cubic lattice is represented by the Hamiltonian
\begin{equation}
 \mathcal{H} = -t\sum_{<ij>}(\hat{a}^{\dagger}_i\hat{a}^{\phantom\dagger}_j + \mathrm{h.c.}) + V\sum_{<ij>}\hat{n_i}\hat{n_j} - \mu\sum_{i}\hat{n}_i.
\end{equation}
The $\hat{a}, \hat{a}^{\dagger}$ annihilation and creation field operators satisfy the usual bosonic commutation relations for different sites and fermionic anticommutation relations for the same site, thereby preventing double occupancy; $\hat{n} = \hat{a}^{\dagger}\hat{a}^{\phantom\dagger}$ is the boson number operator. The interaction $V$ and hopping $t$ are between NN sites, and the chemical potential $\mu$ controls the particle density. The spinless bosons can be mapped to an anisotropic Heisenberg spin-1/2 model with the following equivalence relations: $\hat{a}^{\dagger}_i \leftrightarrow S^{+}_i, \hat{a}_i \leftrightarrow S^{-}_i$ and $\hat{a}^{\dagger}_i\hat{a}^{\phantom\dagger}_i \leftrightarrow S^{z}_i + 1/2$, for any lattice site $i$.\\

\section{Spin wave analysis}
The mean field and spin wave analyses proceed in the familiar textbook way\cite{Ma, Auerbach}. In the subsequent formulation, we closely follow the notation of Ref. [5].\\
Upon changing to the spin representation, the Hamiltonian in Eq. (1) becomes
\begin{eqnarray}
 \mathcal{H} &=& -t\sum_{<ij>}(\hat{S}^{+}_i\hat{S}^{-}_j + \mathrm{h.c.}) + V\sum_{<ij>}\hat{S}^{z}_i\hat{S}^z_j \nonumber \\ &-& (\mu - \frac{zV}{2})\sum_{i}\hat{S}^{z}_i - \frac{1}{2}(\mu - zV/4)N,
\end{eqnarray}
where $z$ is the number of nearest neighbours and $N$ is the number of lattice points. For cubic lattices in $d$-dimensions, $z = 2d$.\\
Consider the mean-field ground state $|\psi\rangle = \prod_{i=1}^{N} (\sin{(\theta /2)} + \cos{(\theta /2)}\hat{a}^{\dagger}_i)|0\rangle$, where $\theta$ is the angle made by the (pseudo)spins with respect to the $z$-axis. We choose, without loss of generality, an ordering such that the $xz$ plane contains all the spins. Minimizing the grand canonical potential per site $\Omega = \langle \psi|\mathcal{H}|\psi \rangle$ with respect to the angle $\theta$, we obtain, for the mean field values, 
\[
 \Omega_{MF} = \rho_{MF}(\frac{z}{2}V\rho_{MF} - \mu + zt\rho_{MF} - zt),
\]
where the mean field particle density 
\[
\rho_{MF} \equiv \cos^{2}{(\theta/2)} = \cfrac{\mu + zt}{z(2t + V)}.
\]
The energy density is given by $E_{MF} = \Omega_{MF} + \mu \rho_{MF}$.\\
Spin wave corrections can be included by twisting the $z$-axis to align with the direction of MF magnetization:
\begin{eqnarray}
 & &\hat{S}^{x}_i = \tilde{\hat{S}}^{x}_i\cos{(\theta)} + \tilde{\hat{S}}^{z}_i\sin{(\theta)} \nonumber \\
 & &\hat{S}^{y}_i = \tilde{\hat{S}}^{y}_i \nonumber \\
 & &\hat{S}^{z}_i = -\tilde{\hat{S}}^{x}_i\sin{(\theta)} + \tilde{\hat{S}}^{z}_i\cos{(\theta)}.\nonumber \\
\end{eqnarray}
In this rotated frame, the excitations are bosonic in nature because the spin-flips change the $\tilde{S}^{z}$ projected spin values by $\pm 1$. This requires a bosonic representation using the Holstein-Primakoff transformations for hardcore bosons
\begin{eqnarray}
 & &\tilde{\hat{S}}^{x}_i = \frac{1}{2}(b^{\dagger}_{i} + b^{\phantom\dagger}_i) \nonumber \\
& &\tilde{\hat{S}}^{y}_i = \frac{1}{2i}(b^{\dagger}_{i} - b^{\phantom\dagger}_i) \nonumber \\
& &\tilde{\hat{S}}^{z}_i = \frac{1}{2} - b^{\dagger}_{i}b^{\phantom\dagger}_i.
\end{eqnarray}
\begin{figure*}[t!]
\centering
    \subfigure[ ]{
    \label{Energy}
    \includegraphics*[width=8cm]{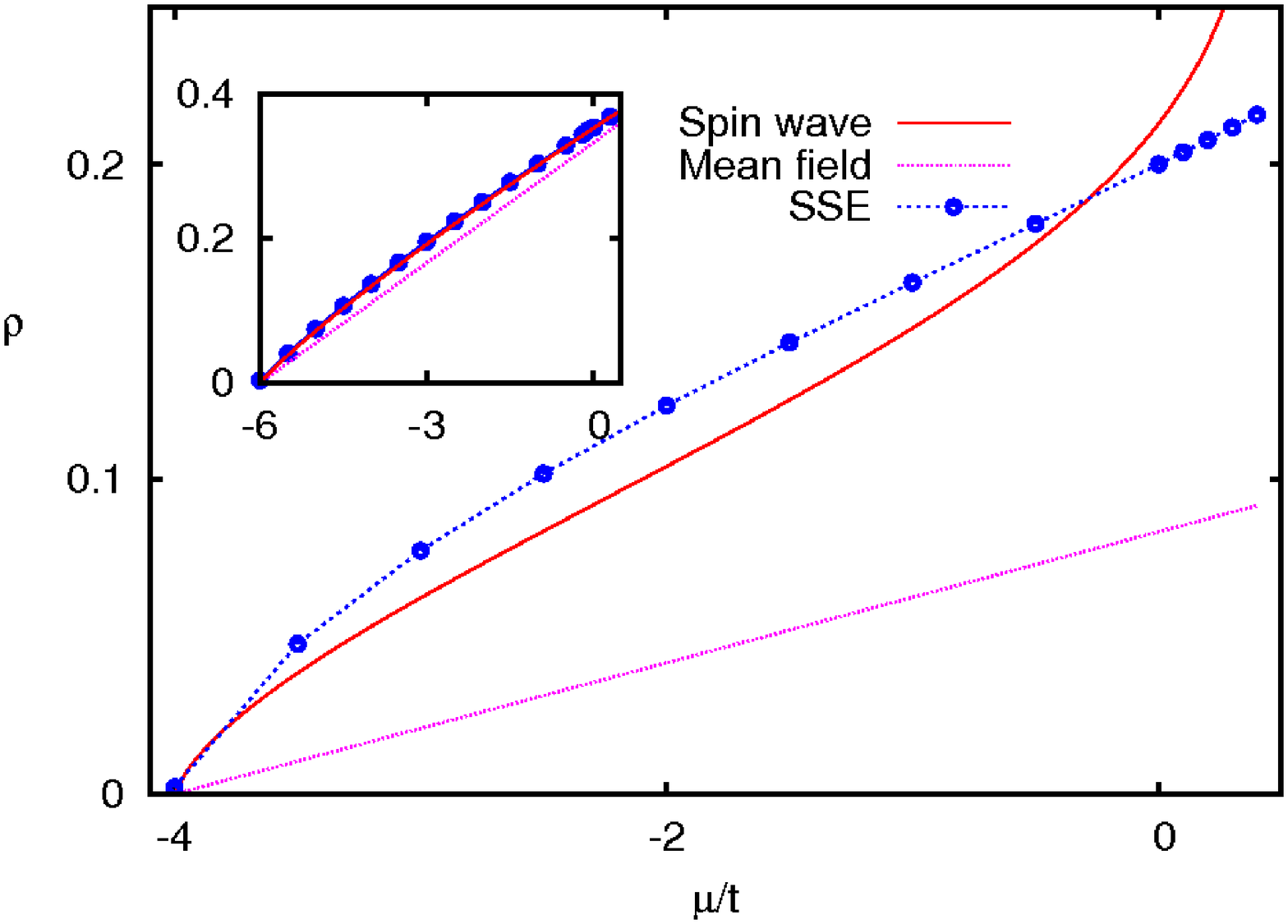}
}
    \subfigure[ ]{
    \label{Condensate}
    \includegraphics[width=8cm]{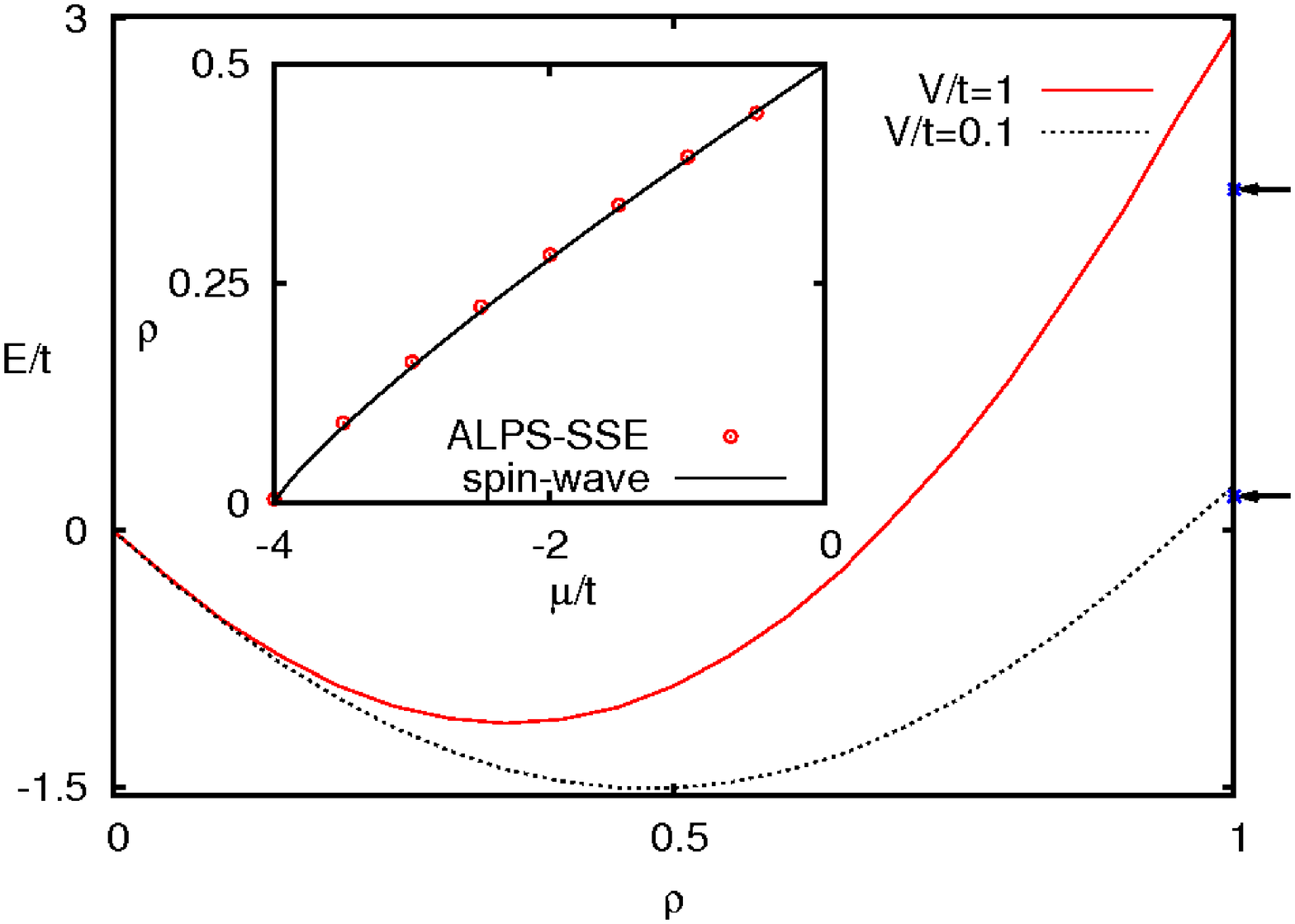}
}
    \caption{(Colour online) (a) Mean field particle density on the square lattice compared with spin-wave results and implementation of Stochastic Series Expansion using the ALPS library\cite{ALPS} at $V/t = 10$. Close to the first-order transition to the checker-board solid, the SWT does not converge. Inset shows similar comparisons of boson density for the simple cubic lattice at $V/t = 1$. (b) Energy density for the simple cubic lattice in the full boson density range, with arrows indicating MF values at $\rho = 1$. Inset shows the boson density computed using spin-wave and ALPS's SSE\cite{ALPS} on the square lattice at $V=0$, verifying the result of Ref. [5].} 
\end{figure*}
With this substitution the spin wave Hamiltonian, up to quadratic terms, is readily shown to be
\begin{widetext}
 \begin{eqnarray}
   \mathcal{H}_{SW} &=& \sum_{<ij>}\{(b^{\dagger}_ib^{\dagger}_j + b^{\phantom\dagger}_ib^{\phantom\dagger}_i)(\frac{t}{2} +  \frac{V}{4})\sin^2{(\theta)} + (b^{\dagger}_ib^{\phantom\dagger}_j + b^{\phantom\dagger}_ib^{\dagger}_j)(-\frac{t}{2}(1 + \cos^2{(\theta)}) + \frac{V}{4}\sin^2{(\theta)})\} + \nonumber \\
& &\sum_{i}b^{\dagger}_ib^{\phantom\dagger}_i[zt\sin^2{(\theta)} - \frac{zV}{2}\cos^2{(\theta)} + (\mu - \frac{zV}{2})\cos{(\theta)}] + N[(\frac{zV}{4} - \frac{\mu}{2})\cos{(\theta)} + \frac{z}{4}(\frac{V}{2}\cos^2{(\theta)} -t\sin^2{(\theta)}) \nonumber \\
& &+ \frac{zV/4-\mu}{2}] + \sum_i[\frac{\mu-zV/2}{2}\sin{(\theta)} - \frac{z}{4}(2t+V)\cos{(\theta)}\sin{(\theta)}](b^{\dagger}_i + b^{\phantom\dagger}_i).
 \end{eqnarray}
\end{widetext}
So that the particle number is conserved in the new bosonic system, we require that each of the last summands disappear. This gives the spin-wave condition for $\theta$ to be 
\[
 \cos{(\theta)} = \frac{2\mu - zV}{z(2t + V)}.
\]
 This was the same condition previously obtained without the spin waves as well, noted as well in the non-interacting case\cite{Bernardet}. We mention in passing that we can, from the above, obtain a $d$-dimensional analog of the Pokrovsky-Talapov line\cite{Pokrovskii} separating the massless and the ferromagnetic phases in the one-dimensional anisotropic Heisenberg model in a longitudinal field by equating $\cos{(\theta)} = 1$. This gives the second order phase separating line between the MI and SF phases to be
\begin{equation}
 (\mu)^{PT}_c = z(t + V).
\end{equation}
The spin wave Hamiltonian retains the same form as in Ref. [5] except for the coefficients of the operators. Therefore the Bogoliubov transformation 
\begin{eqnarray}
 & &b_{\vec{\mathbf{k}}} = u_{\vec{\mathbf{k}}}\alpha_{\vec{\mathbf{k}}} - v_{\vec{\mathbf{k}}}\alpha^{\dagger}_{-\vec{\mathbf{k}}} \nonumber \\
& &b^{\dagger}_{\vec{\mathbf{k}}} = u_{\vec{\mathbf{k}}}\alpha^{\dagger}_{\vec{\mathbf{k}}} - v_{\vec{\mathbf{k}}}\alpha_{-\vec{\mathbf{k}}}.
\end{eqnarray}
to separate the MF ground state and the excited states above it - represented by the $\alpha$ bosons - immediately simplifies to Eq. (22) in Ref. [5] but with different expressions for the coefficients i.e.
\begin{eqnarray}
 \mathcal{H}_{SW} &=& \mathcal{H}_{MF} + \sum_{\vec{\mathbf{k}}}(\sqrt{A^2_{\vec{\mathbf{k}}} - B^2_{\vec{\mathbf{k}}}} - A_{\vec{\mathbf{k}}}) \nonumber \\&+& \sum_{\vec{\mathbf{k}}}\sqrt{A^2_{\vec{\mathbf{k}}} - B^2_{\vec{\mathbf{k}}}}(\alpha^{\dagger}_{\vec{\mathbf{k}}}\alpha^{\phantom\dagger}_{\vec{\mathbf{k}}} + \alpha^{\dagger}_{-\vec{\mathbf{k}}}\alpha^{\phantom\dagger}_{-\vec{\mathbf{k}}}).
\end{eqnarray}
The sum over $\vec{\mathbf{k}}$ is over the first Brillouin zone and the mean field Hamiltonian corresponds to the mean field grand canonical potential $\Omega_{MF}$. The variables $A_{\vec{\mathbf{k}}}, B_{\vec{\mathbf{k}}}$ are here given by
\begin{align*}
 A_{\vec{\mathbf{k}}} &= \frac{zt}{2} + \frac{\gamma_{\vec{\mathbf{k}}}}{4}[(V - 2t) - \frac{(2\mu - zV)^2}{z^2(2t + V)}],\\
B_{\vec{\mathbf{k}}} &= \frac{\gamma_{\vec{\mathbf{k}}}}{z^2(2t + V)}[tz^2(t+V) + \mu(zV-\mu)], \\
\gamma_{\vec{\mathbf{k}}} &= \sum^{\frac{z}{2}}_{i=1}\cos{(k_i)}.
\end{align*}
Here the lattice constant was taken to be one and gauge invariance in the $\alpha$ operators is ensured by choosing
\[
 u^2_{\vec{\mathbf{k}}} (v^2_{\vec{\mathbf{k}}}) = \frac{1}{2}(\frac{A_{\vec{\mathbf{k}}}}{\sqrt{A^2_{\vec{\mathbf{k}}} - B^2_{\vec{\mathbf{k}}}}} \pm 1).
\]

\section{Thermodynamic quantities}
\begin{figure*}[t!]
\centering
    \subfigure[ ]{
    \label{DensityJump}
    \includegraphics*[width=8cm]{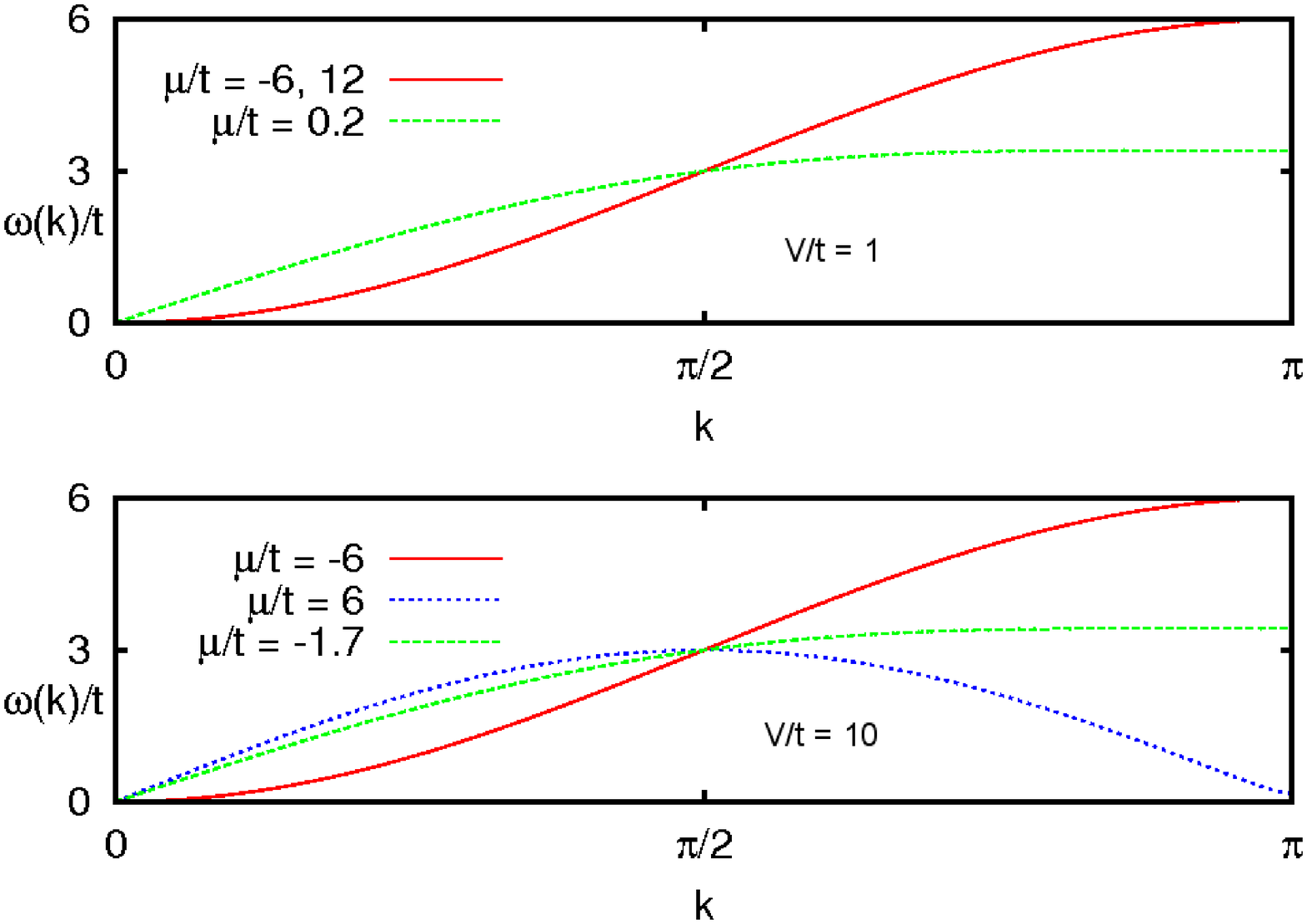}
}
    \subfigure[ ]{
    \label{PhaseDiagram}
    \includegraphics[width=8cm]{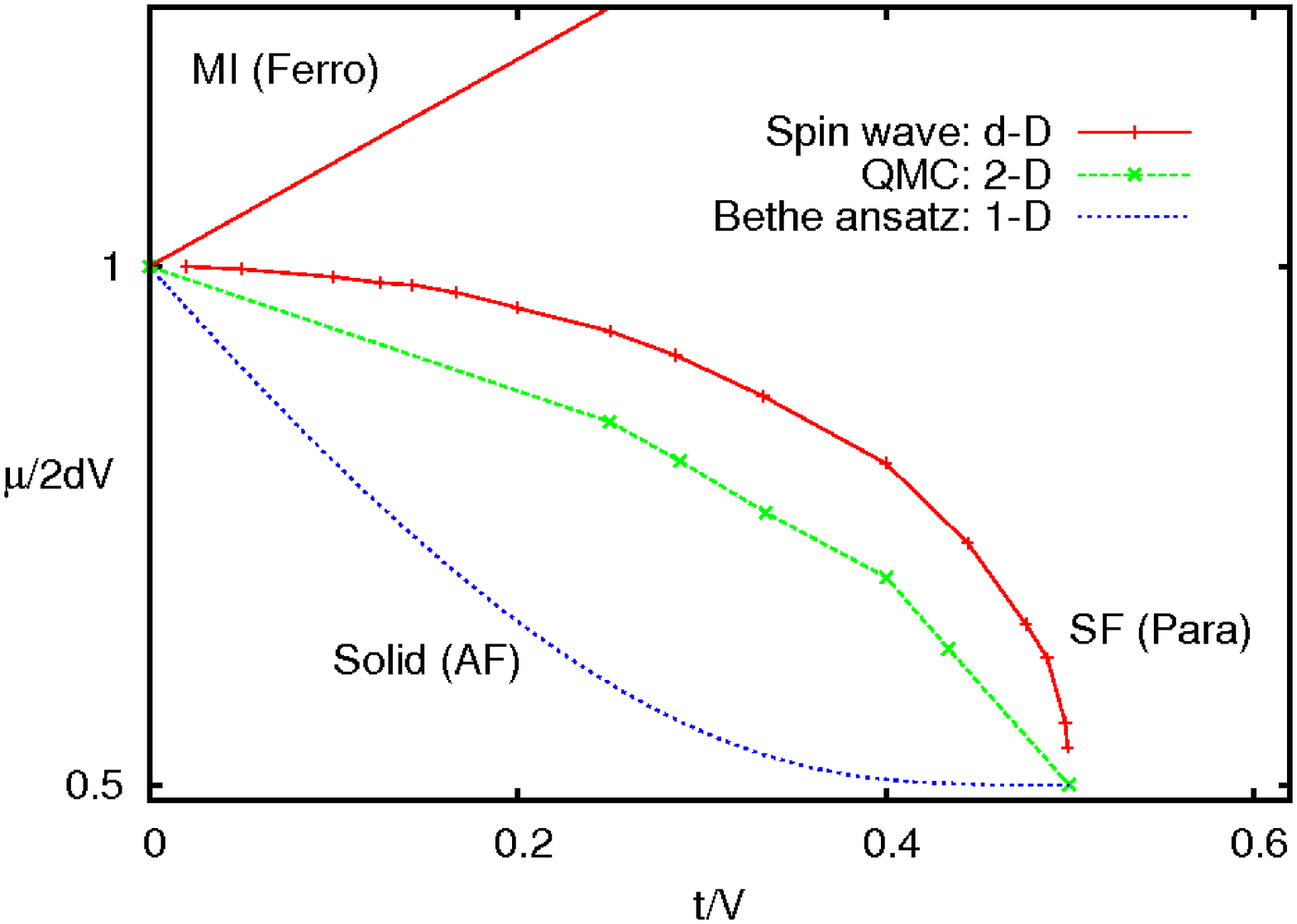}
}
    \caption{(Colour online) (a) Quasiparticle - $\alpha$ bosons - dispersion for 2 interaction strengths along $k_x = k_y = k_z$ in the 3D simple cubic lattice. As explained in the text, ferromagnetic and antiferromagnetic ordering is readily seen from behaviour near the zone edge or centre. (b) SWT phase diagram compared with other known results like QMC\cite{Hebert} on a square lattice and Bethe ansatz on a linear chain. Naturally, we expect the SWT to be more accurate in higher dimensions.} 
\end{figure*}
Using the above results, we can generalize the results in Ref. [5] to a cubic lattice in arbitrary dimensions. The average particle density, including the spin waves but calculated within the MF state, is given by
\begin{eqnarray}
 \rho_{SW} &=& -\frac{\partial{\Omega_{SW}}}{\partial{\mu}} = \rho_{MF} + \nonumber \\& & \mu \frac{2\mu-zV}{z^2N(2t + V)}\sum_{\vec{\mathbf{k}}}\gamma_{\vec{\mathbf{k}}}(\frac{A_{\vec{\mathbf{k}}} - B_{\vec{\mathbf{k}}}}{\sqrt{A^2_{\vec{\mathbf{k}}} - B^2_{\vec{\mathbf{k}}}}} - 1),
\end{eqnarray}
where $\Omega_{SW}$ is given by the expectation value of the first line of Eq. (8), whereas the coefficient $\sqrt{A^2_{\vec{\mathbf{k}}} - B^2_{\vec{\mathbf{k}}}}$ in the second line represents the dispersion of the excitations. In Fig. 1(a) we compare the boson density calculated using SWT, with results of Quantum Monte Carlo (SSE implemented using the ALPS library\cite{ALPS}) simulations and MF analysis on the square lattice for $V/t = 10$. In the SSE simulation we used $\beta \equiv 1/T = 40$ on a 25x25 system using half a million sweeps and a thermalization of ten thousand; the error bars are not shown (because they are only as large as the symbol size) and simulations were done for various system sizes ($L = 25, 12, 6, 4$), sweeps, thermalization and inverse-temperatures ($\beta = 40, 20, 10$) to check for convergence. As can be seen SWT is an improvement over the MF results but compares only approximately with SSE, with the differences increasing towards the phase boundary; the comparison is not as good as with the non-interacting system in Ref. [5], which we have verified as shown in the inset of Fig. 1(b). For one dimension higher i.e. simple cubic lattice, and smaller $V/t$, spin wave analysis compares well with an SSE simulation using the ALPS library as shown in the inset of Fig. 1(a) for $V/t = 1$; the system size used was 5x5x5 with the same thermalization, sweeps and temperature as mentioned above.\\

The energy density in the SW approximation is obtained as before within mean field i.e. $E_{SW} = \Omega_{SW} + \mu \rho_{SW}$.
Finally, the functional expressions for the expectation value of the condensate fraction $\rho_0$ - particle density in the zero momentum sector - within MF and SW remain the same as in Ref. [5] in the absence of NN interactions:
\begin{eqnarray}
 \rho_{0,MF} = \rho_{MF}(1 - \rho_{MF}), \nonumber \\
\rho_{0,SW} = \rho (1 - \rho) - \frac{1}{N}\sum_{\vec{\mathbf{k}} \neq \vec{\mathbf{0}}}v^2_{\vec{\mathbf{k}}}.
\end{eqnarray}
The energy density in the simple cubic lattice for 2 different interaction strengths $V$ are computed as functions of the SW particle density by Legendre transforming from the $\mu$ variable and shown in Fig. 1(b). The condition $E(\rho) = E(1-\rho)$ is not satisfied at a finite $V$ because it is clear that larger particle densities contribute more to the intersite repulsive energy. From the figure, the usual expectations are borne out: higher intersite repulsion adds more SW contributions to the MF value. That is, lower the interaction, better is the MF description. This may be seen by the distance between the location of the arrows (MF values) and the corresponding SW values for two different interaction strengths. It is worth noting from Eq. (10) that the MF condensate fraction depends only on the particle density and not on the interaction strength; whereas the larger the $V$ value, the more will Bose-Einstein condensation be suppressed within SWT.

\section{Excitations}

As seen in the previous section, the excitation spectrum of the $\alpha$ bosons is given by $\omega(\vec{\mathbf{k}}) = \sqrt{A^2_{\vec{\mathbf{k}}} - B^2_{\vec{\mathbf{k}}}}$. The quantities $A_{\vec{\mathbf{k}}}, B_{\vec{\mathbf{k}}}$ still depend on $\mu, V$ but the dependence is notationally suppressed. The dispersion is calculated along the line $k_x = k_y = k_z$ for a simple cubic lattice and plotted in Fig. 2(a) for two different interaction strengths. \\
Far away from the expected checkerboard solid phase, antiferromagnetic (AF) ordering is not expected for any particle density. This is indeed true and checked for $V/t = 1$ for all range of densities between the empty and full phases. That is, the dispersion in this case is always gapped at $\vec{\mathbf{k}} = (\pi, \pi, \pi)$. The dispersion at the boundaries of these two ferromagnetic phases - $\mu/V = -6, 12$ - show $\lim_{|\vec{\mathbf{k}}| \to 0}\omega (\vec{\mathbf{k}}) \propto |\vec{\mathbf{k}}|^2$ (upper panel of Fig. 2(a)) as expected for such ordering. The gap to magnon excitations in the MI phase is of course just given by $\Delta = (\mu - (\mu)^{PT}_c)/2$.\\
In the strong coupling limit - $V/t = 10$ - antiferromagnetic ordering sets in as the checkerboard solid is approached. This is seen in the lower panel of Fig. 2(a), where for $\mu/V = 0.6$, 
\[
    \omega (\vec{\mathbf{k}}) \propto 
\begin{cases}
    |\vec{\mathbf{k}}|,&  |\vec{\mathbf{k}}|\approx 0\\
    |\vec{\mathbf{k}} - (\pi,\pi,\pi)|,              & |\vec{\mathbf{k}}|\approx (\pi,\pi,\pi).
\end{cases}
\]

Finally, although the SW ground state phase diagram has already been fleshed out in detail\cite{Pich}, our purpose here is merely to compare it with other known results. The SW construction proceeds in the usual way: for every interaction strength, the particle density is calculated for all chemical potentials. $\rho$ will have a discontinuity at two values of $\mu$ corresponding to the first order phase transition between the solid and superfluid. These will define the boundaries of the expected checkerboard solid lobe. Moreover, it must be recalled that within our SWT, only collinear phases are describable because we chose $\theta_i = \theta$. Therefore, we do not expect convergence in the solid phase anyway. With these assumptions, the SWT phase diagram is constructed for the HCB model on any $d$-dimensional cubic lattice; the separation between the massless and the ferromagnetic phase was already given by Eq. (6).\\
This is compared to QMC calculations in two dimensions\cite{Hebert} and the Bethe ansatz solution in one dimension. The latter was solved for the anisotropic Heisenberg model on a chain\cite{Yang}; the critical line $(\frac{\mu}{V})_c$ separating the massless and antiferromagnetic phase is given by
\begin{equation}
 \frac{V}{2t}[(\frac{\mu}{V})_c - 1] = \sinh{(\lambda)}\sum^{\infty}_{n = -\infty}(-1)^n\cosh^{-1}{(n\lambda)},
\end{equation}
where $\lambda = \cosh^{-1}{(\frac{V}{2t})}$.
The Pokrovsky-Talapov line remains the same. The comparisons between Bethe ansatz in 1D, QMC in 2D and SWT in $d$-D is shown in Fig. 2(b).\\

To summarize, we have generalized the SWT for cubic lattices in arbitrary dimensions for a hardcore boson model with nearest neighbour interaction. Based on earlier work\cite{Bernardet,Hen}, where in Ref. [5] it was suggested that such an SWT should give reasonable results for the interacting system as well, we find that the comparison with SSE is not as good as it is with the non-interacting case for the square lattice and strongly repelling nearest neighbour bosons. Although we did not start with a bond-ordered phase, as suggested in Ref. [5], we can compute observables like energy density and condensate fraction as long as the system is not within the checkerboard solid phase: the collinear MF states should, in principle, describe the SF and MI phases. These quantities are evaluated for the simple cubic lattice and compared with the mean field values. In a simple cubic lattice and away from the strongly-repulsive regime, we find good comparison for the particle-density between results of spin-wave and SSE simulations.

\acknowledgments

We thank G.~G. Batrouni for extensive clarifications. Discussions with H. Frahm, V. Rittenberg and S. G\"{u}rtler are acknowledged. VKV thanks the Bonn-Cologne Graduate School for funding within the Deutsche Forschungsgemainschaft.

\bibliographystyle{unsrt}
\bibliography{Ref}

\end{document}